\documentclass[twocolumn]{aastex62}
\usepackage[utf8]{inputenc}
\usepackage{amsmath, amssymb, amsthm, mathtools}
\usepackage{wasysym}
\usepackage{color}
\usepackage{hyperref}
\usepackage{xspace}
\usepackage{xcolor}
\usepackage{natbib}

\usepackage{soul} 

\received{2019 October 21}
\revised{2020 May 29}
\accepted{2020 June 8}
\submitjournal{The Astrophysical Journal}

\shorttitle{Shaping the outer disc of NGC 4565 via accretion}
\shortauthors{Gilhuly et al.}

\begin{document}

\title{The Dragonfly Edge-on Galaxies Survey: Shaping the outer disc of NGC 4565 via accretion}

\correspondingauthor{Colleen Gilhuly}
\email{gilhuly@astro.utoronto.ca}

\author{Colleen Gilhuly}
\affil{Department of Astronomy \& Astrophysics, University of Toronto, 50 St. George Street, Toronto, ON M5S 3H4, Canada}
\author{David Hendel}
\affil{Department of Astronomy \& Astrophysics, University of Toronto, 50 St. George Street, Toronto, ON M5S 3H4, Canada}
\author{Allison Merritt}
\affil{Max-Planck-Institut f{\"u}r Astronomie, K{\"u}nigstuhl 17, D-69117 Heidelberg, Germany}
\author{Roberto Abraham}
\affil{Department of Astronomy \& Astrophysics, University of Toronto, 50 St. George Street, Toronto, ON M5S 3H4, Canada}
\author{Shany Danieli}
\affil{Astronomy Department, Yale University, 52 Hillhouse Ave, New Haven, CT 06511, USA}
\affil{Physics Department, Yale University, 52 Hillhouse Ave, New Haven,
CT 06511, USA}
\author{Deborah Lokhorst}
\affil{Department of Astronomy \& Astrophysics, University of Toronto, 50 St. George Street, Toronto, ON M5S 3H4, Canada}
\author{Qing Liu}
\affil{Department of Astronomy \& Astrophysics, University of Toronto, 50 St. George Street, Toronto, ON M5S 3H4, Canada}
\author{Pieter van Dokkum}
\affil{Astronomy Department, Yale University, 52 Hillhouse Ave, New Haven, CT 06511, USA}
\author{Charlie Conroy}
\affil{Harvard-Smithsonian Center for Astrophysics, 60 Garden Street, Cambridge, MA 02138, USA}
\author{Johnny Greco}
\affil{Center for Cosmology and Astroparticle Physics (CCAPP), The Ohio State University, Columbus, OH 43210, USA}

\begin{abstract}

We present deep $g$- and $r$-band imaging of the well-known edge-on galaxy NGC 4565 (the ``Needle Galaxy''), observed as part of the Dragonfly Edge-on Galaxies Survey. The $3\sigma$ local surface brightness contrast limit on 10 arcsec scales is $28.616 \pm 0.005$ mag/arcsec$^{2}$ for the $r$-band image and $28.936 \pm 0.005$ mag/arcsec$^{2}$ for the $g$-band image. We trace the galaxy's starlight in narrow slice profiles spanning over 90 kpc along the major axis (with bin sizes ranging from $1.7 \times 0.5$ kpc to $1.7 \times 7.8$ kpc) to surface brightnesses below 29~mag~arcsec$^{-2}$. We confirm the previously observed asymmetric disc truncation in NGC 4565. In addition, the sharp northwest truncation turns over to a shallower component that coincides with a fan-like feature seen to wrap around the northwest disc limb. We propose that the fan may be a tidal ribbon and qualitatively replicate the fan with simple simulations, although alternative explanations of the fan and the disc's asymmetry are shown to be possible. In any case, we conclude that there is strong evidence for accretion-based outer disk growth in NGC 4565. 

\end{abstract}


\section{Introduction} \label{intro}

Our understanding of the formation and growth of spiral galaxies in a hierarchical, dark matter-dominated universe is constantly evolving. \cite{fallefstathiou80} developed a model to describe the formation of thin, rotating disc galaxies by the collapse of slowly rotating gas within a massive dark matter halo. Stellar feedback was found to be crucial for the formation of realistic discs in simulations \citep{whitefrenk91, mo98}. Spiral galaxies were thought to have avoided any recent major mergers, as otherwise they would have been transformed into elliptical galaxies \citep[and many others]{negropontewhite83}. 

Minor mergers initially did not fit comfortably into this picture. Dissipationless N-body simulations showed that minor mergers cause thin discs to spread, warp, and flare, coming to resemble thick discs or an S0/Sa galaxy \citep{quinn93, walker96}. Cosmological simulations and observations both showed minor mergers to be very common, which raised questions of how so many thin discs in spiral galaxies were able to survive to the present day \citep[and references therein]{hopkins09}. 

Improvements in simulations of mergers, particularly the inclusion of gas, resolved this tension. 
It was first shown that gas-rich major mergers can result in a disc-dominated remnant if star formation is suppressed during the interaction (by ISM pressurization) such that the system is gas-dominated at the moment of coalescence \citep{springelhernquist05, robertson06}. Such mergers could plausibly contribute to the assembly of today's spirals at high redshift, where progenitors are more likely to have high gas fractions. When gas is available in a spiral galaxy, the degree of disc heating from minor mergers is greatly reduced and a recognizeable thin disc survives the encounter \citep{hopkins09, scannapieco09, moster10}. More realistic orbits in improved dissipationless N-body simulations also led to decreased disc heating, showing that minor mergers are not catastrophic for spiral galaxies \citep{hopkins08, kazantzidis08}.

Thin discs may be able to \emph{survive} minor mergers, but is it also possible to \emph{grow} these discs with accretion? The accretion of satellite galaxies is often thought of in terms of bulge growth and stellar halo assembly \citep{helmiwhite99, bullockjohnson05, abadi06}. However, recent work shows that minor mergers do lead to disc growth and indeed may be necessary to explain the growth of discs since $z \sim 2$ as star formation and stellar migration alone are likely insufficient \citep{bluck12, ownsworth12, sachdeva15}. These claims are supported by cosmological hydrodynamic simulations, where stars that formed in satellite galaxies (``ex-situ'') are found throughout discs at the present day \citep{pillepich15, rodriguezgomez16}. Minor mergers are also an effective way to explain the size growth of early type galaxies over this period as major mergers do not increase size efficiently  \citep{bezanson09, naab09}.

If ongoing minor mergers play an important role in the growth of discs, there should be some hint of the origin of recently accreted material. The many streams surrounding the Milky Way plainly reveal its extensive and ongoing merger history \citep{belokurov06, helmi17, malhan18}. The age-velocity dispersion relation of stars in the solar neighbourhood \citep{quillen01} and disruptions in stellar phase-space \citep{gomez12} may show the impact of these past mergers on the disc. These high-dimensional approaches are not feasible for galaxies other than our own. 

For relatively nearby galaxies, resolved stellar photometry can be used to assess changes in the age or metallicity of stars across the disc \citep[eg.][]{radburnsmith12, streich16}. Otherwise, one can examine the morphology and structure of galaxies in wide field broadband imaging to search for signatures of recent minor mergers. It has been suggested that broken exponential profiles with a larger outer scalelength (Type III) arise due to minor mergers \citep{younger07}. However, there are many other possible explanations: smooth gas accretion combined with radial migration \citep{minchev12}, a superposition of thin and thick discs \citep{comeron12}, tidal interactions, galaxy harassment \citep{watkins19}, or simply the onset of the stellar halo-dominated regime. An upbending outer profile does not necessarily imply outer disc assembly via minor mergers.  

Stellar streams near the disc plane, outer warps, or other disturbances in the outskirts of a disc  are more convincing evidence of recent accretion events than profile inflections alone. Dedicated, deep integrated light observations are required in order to follow the outer disc to its farthest reaches to search for such clues. This approach is complicated or precluded by scattered light within conventional reflecting telescopes. The Dragonfly Telephoto Array has been specifically designed to minimize scattered light  and has been successfully used to study a variety of very low surface brightness structures, including low surface brightness dwarfs \citep{merritt14, vandokkum15}, stellar halos \citep{merritt16} and extended stellar discs \citep{zhang18}.

In this paper, we present deep observations of the edge-on galaxy NGC 4565 with Dragonfly. In Section~\ref{obs} we describe our observations and data reduction procedure. Section~\ref{source_sub} focuses on the important areas of point-spread function characterization and source subtraction. Our analysis is presented in Section~\ref{analysis}, and discussed in Section~\ref{discussion}. We summarize this paper in Section~\ref{summary}.

Throughout, we adopt a distance of 12.7 Mpc to NGC 4565. This is the average of the two tip of the red-giant branch distances tabulated on the NASA/IPAC Extragalactic Database \citep[11.9 and 13.5 Mpc;][respectively]{radburnsmith11, tully13}. Jupyter notebooks containing the code used to produce the profiles and the figures in this paper are available on GitHub\footnote{https://github.com/cgilhuly/papers/tree/master/2019/NGC4565\_disc}.

\section{Observations and data reduction} \label{obs}

The Dragonfly Telephoto Array \citep{abraham14} is a robotic mosaic telescope currently consisting of 48 Canon 400mm $f$/2.8L IS II USM telephoto lenses. State-of-the-art anti-reflective coatings on optical surfaces and the absence of obstructions in the optical path lead to significantly less scattered light than in typical reflecting telescopes. This makes Dragonfly well-suited for observations of low surface brightness objects and features. A generous field of view ($2.6^\circ \times 1.9^\circ$) avoids irregularities where individual pointings are stitched together and enables confident detection of faint features unhindered by image artifacts. In its current configuration, Dragonfly is equivalent to a 0.99-m $f$/0.4 refracting telescope. 

The Dragonfly Edge-on Galaxies Survey (DEGS) is an imaging campaign targeting nearby edge-on spiral galaxies (C. Gilhuly et al., in prep). Galaxies were selected using criteria of distance, absolute magnitude, inclination, and absence of Galactic cirrus. Dragonfly's large field of view and excellent sensitivity to low surface brightness structures enables the exploration of galactic disc outskirts and stellar halos, and the edge-on orientation of the sample allows a cleaner separation of disc and halo components.

NGC 4565 was the first galaxy in the DEGS sample to be observed, and it was imaged for significantly longer than typical DEGS targets. A more ``typical''  DEGS galaxy observation is that of NGC 5907, presented in \cite{vandokkum19}. We observed NGC 4565 over 64 nights from Spring 2015 to Summer 2017. The majority of our data originate from 42 nights in 2017, after Dragonfly was upgraded to a 48-lens configuration. In total, we collected 29473 10-minute exposures (equivalent to 102.3 hours with the entire array). This is Dragonfly's deepest field to date. Half of the array is equipped with SDSS $g$ filters, and the other half with $r$ filters. These filters are similar to SDSS filters; see Figure~2 of \cite{abraham14} for details. Dithers of  35 arcmin and slight misalignment of the lenses were used to improve control over sky systematics.

We produced coadded images of the NGC 4565 field using the Dragonfly reduction pipeline. A detailed description of this software can be found in Section~3 of \cite{danieli20}. One key strength of the pipeline is the two-stage sky subtraction applied to individual frames. The sky is first modelled using source masks generated from each frame. The intermediate sky-subtracted frames are then median-combined to produce an intermediate mosaic. This mosaic is then used to define a more robust source mask due to its greater depth, enabling better sky models and ultimately a more uniform final image.

Another relevant strength of the pipeline is its ability to reject poor quality frames obtained under conditions where very light cloud contamination or high atmosphere aerosols enhances the wings of the point-spread function (PSF). This subtle effect is not always visible to the eye, but we are able to exclude such data by rejecting deviant zeropoints during photometric calibration. This data pruning technique allows Dragonfly to observe in all conditions, making use of brief photometric windows without contaminating the final images. 

The final images of the NGC 4565 field contain 4448 $r$-band frames and 4416 $g$-band frames, equivalent to 30.8 hours with the entire array. A low conversion rate from raw exposure time to final images is expected given our data pruning technique, described above. These images are available for download at \url{https://www.dragonflytelescope.org/data-access.html}.

\section{PSF management and source subtraction} \label{source_sub}

Scattered light can lead to overestimations of thick disc and stellar halo components of galaxies, and great change in their outer colour profiles \citep{dejong08, sandin14, sandin15}. Therefore, when studying low surface brightness regions in galaxies, particularly edge-on galaxies, it is crucial to characterize the PSF to large radii. The ideal approach is to perform PSF deconvolution on an image, but this is a significant undertaking \citep{karabal17}. Often the more tractable approach is to fit the galaxy with a two-dimensional model, and then to compare model images produced with and without PSF convolution. This provides an estimate of the contribution by scattered light in all regions of the galaxy. This technique has been adopted in numerous recent low surface brightness studies \citep{trujillofliri16,borlaff17,peters17,comeron18,martinezlombilla19,mlknapen19}, which themselves demonstrate the importance of wide angle PSF characterization and correction.

While the unique design of Dragonfly reduces the amount of scattered light at large angles relative to traditional telescopes, it remains important to estimate the impact that scattered light may have on galaxies and other features in our images. 2D Bayesian modelling of scattered light was used to characterize the wide-angle PSF in our final coadded images (Liu et al., in prep). The inner 5 arcsec of the PSF was modelled with a Moffat function and the outer PSF was fitted with multiple power laws to a radius of 20 arcmin, and extrapolated to a radius of 33.33 arcmin. The outer portions of our $r$-band PSF was well-modelled by two power laws ($n = 3.44$ from 5 arcsec - 4 arcmin, $n = 3.02$ from 4 arcmin outwards) while the $g$-band PSF required three ($n = 3.44$ from 5 arcsec - 64.6 arcsec, $n = 2.89$ from 64.6 arcsec - 117.5 arcsec, and $n = 2.07$ from 117.5 arcsec outwards). We cannot rule out the possibility that low amplitude variations in the sky on scales of $\sim10$ arcmin may have affected our modelling efforts, and we are continuing to investigate. We refer the interested reader to Liu et al. (in prep) for further details.

Scattered light from stars and compact sources was mitigated using a technique called ``multi-resolution filtering'' (MRF). We refer the interested reader to van Dokkum et al. (2020) for a detailed description of this technique. In brief, sources were identified and segmented by running Source Extractor \citep{bertin96} on a higher-resolution reference image. A model image of the detected sources was then convolved with a kernel constructed to approximately convert from the PSF of the high-resolution image to that of Dragonfly (Eqn. 4 of van Dokkum et al. 2020). The stellar aureoles of stars brighter than 16.5 magnitudes were also modelled to a radius of 33.33 arcmin, using the parametric wide-angle PSF described above. The core PSF (constructed empirically from unsaturated stars) was flux-matched to the wide-angle PSF at a radius of 30 arcsec. Finally, the convolved model image was subtracted from the Dragonfly image. 

For this purpose, we used archival MegaPipe \citep{gwyn08} CFHT images accessed from the Canadian Astronomy Data Centre. The images are $1.23^\circ \times 1.18^\circ$, somewhat off-centre relative to NGC 4565, and are sampled at 0.186 arcsec. The $r'$-band image combines 9 individual 5 minute MegaPrime exposures and has an image quality of 1.09 arcsec. The $g'$-band image combines 5 individual 5 minute exposures and has an image quality of 1.27 arcsec. 


The source subtraction software automatically masked the worst residuals at the core of bright objects. Additional masking was done by hand for extended emission surrounding background galaxies, sources that were not modelled and subtracted (due to low surface brightness or overlap with a much brighter source), and surrounding the brightest stars in the field. In addition to residuals from source subtraction, we masked regions contaminated by background galaxy clusters southeast of NGC 4565 (400d J1236+2550, $z = 0.175$; GMBCG J189.25931+25.84396, $z = 0.308$; WHL J123647.1+255131, $z = 0.1823$) to avoid contamination from the galaxies' outskirts or any intracluster light. 

We estimated the effective depth of the source-subtracted images using \texttt{sbcontrast}, a code bundled with MRF. The code binned the entire image to the desired scale, subtracted a local background from each binned pixel based on the surrounding 8 pixels, and calculated the variation among the binned and background-subtracted pixels. This technique was designed to measure the typical contrast of objects of a given size with their immediate surroundings. The $3\sigma$ surface brightness limit on 10 arcsec scales is $28.616 \pm 0.005$ mag/arcsec$^{2}$ for the $r$-band image and $28.936 \pm 0.005$ mag/arcsec$^{2}$ for the $g$-band image. On scales from 1 arcmin to 4 arcmin, these limits are somewhat fainter ($\sim$29.0 mag/arcsec$^{2}$ for $r$-band, $\sim$29.6 mag/arcsec$^{2}$ for $g$-band).

\section{Analysis} \label{analysis}
 
\subsection{Overview} \label{analysis-overview}
 
\begin{figure*}[tbp]
\plotone{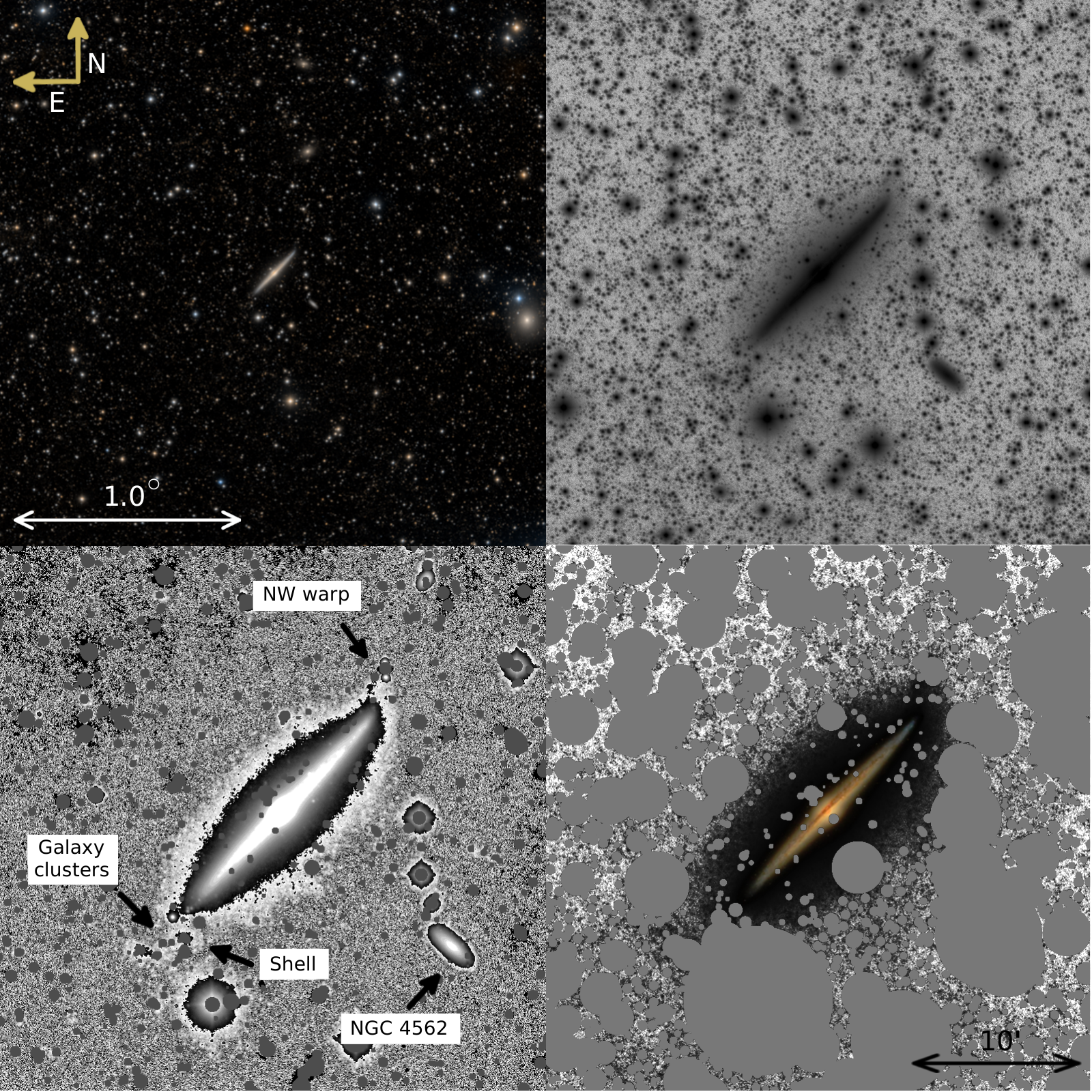}
\caption{Four different views and visualizations of NGC 4565 as imaged by Dragonfly. The top left panel is a false colour composite combining Dragonfly $g$- and $r$-band images, showing a large portion of the final image. The top right panel is a central cut-out of the $r$-band image shown with Gauss histogram equalization scaling. The lower left panel shows the source-subtracted $r$-band image and uses a 4-cycle sawtooth colour palette. The lower right panel shows the source-subtracted $r$-band image (with histogram equalization stretch) where high surface brightness inner regions of NGC 4565 have been replaced with a colour image from the Sloan Digital Sky Survey. Our full source mask is shown in the lower right panel; in general, we show source-subtracted images without our full mask applied. The scale bar in the lower right panel applies to both lower panels.}
\label{fig:quad_images}
\end{figure*}

 \begin{figure}[tb]
\plotone{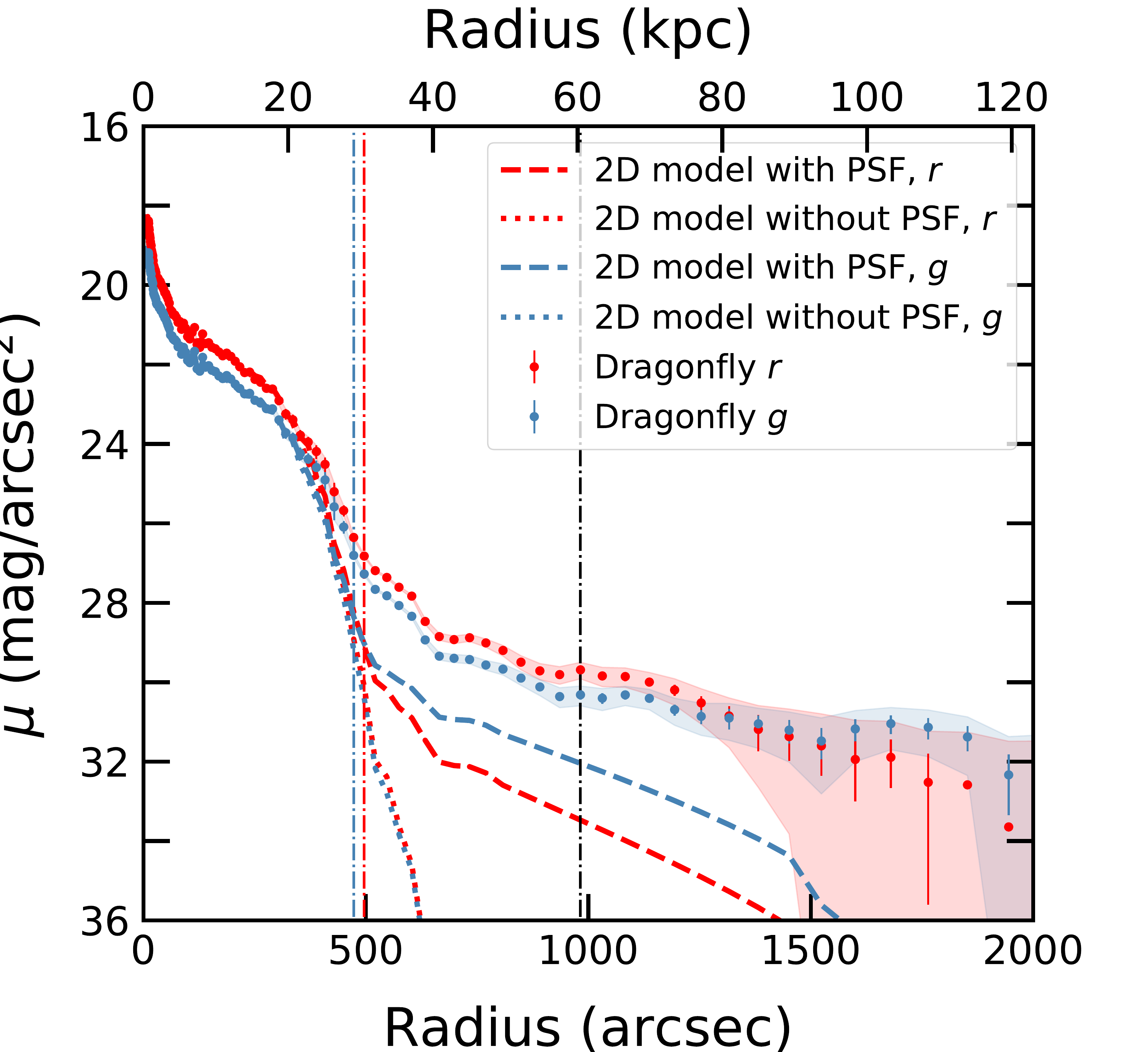}
\caption{ Azimuthally averaged $r$- and $g$-band surface brightness profile of NGC 4565. Error bars depict random errors in each bin, and the shaded envelope also includes systematic uncertainty in the measurement of the sky. The dashed and dotted red (blue) lines indicate the surface brightness profile of the $r$-band ($g$-band) model images generated with and without PSF convolution, respectively. The red (blue) vertical dash-dotted line indicates the radius where the PSF-convolved $r$-band ($g$-band) model profile first becomes at least twice as bright as the model profile without PSF convolution. The black vertical dash-dotted line is an additional outer radius where the contribution of scattered light to both observed profiles is estimated. Both model profiles were extracted using the r-band isophotal solution. }
\label{fig:profiles}
\end{figure}

In Figure~\ref{fig:quad_images}, we present several different views of our reduced and processed images of NGC 4565. The upper left panel shows a large area of the NGC 4565 field in false colour, spanning $2.4^\circ \times 2.4^\circ$. The upper right panel offers a closer look at the galaxy in the $r$-band mosaic. The bottom panels show the source-subtracted $r$-band cut-out. Our source mask is included in the bottom right panel; the ``raw'' source-subtracted image (with automatic masking of core residuals only) is shown in the bottom left panel. For reference, we have replaced the bright regions of NGC 4565 with an SDSS $gri$ colour cut-out in the lower right panel.

Dragonfly is  poorly suited for imaging point sources to great depth due to its relatively large pixels, but it excels in revealing extended low surface brightness features. In the NGC 4565 field, structures with $\mu_r = 27-28$~mag/arcsec$^\textrm{2}$ are clearly visible without any binning or enhancement of the image. Some examples include a large tidal tail around a z = 0.02 galaxy, tidal structures in a cluster at z = 0.175 (as well as extended intracluster light), and features likely associated with the outer disc of NGC 4565 (discussed below; see also Figure~\ref{fig:fan}). These examples are consistent with the estimated $3\sigma$ surface brightness contrast limit of $28.616 \pm 0.005$ mag/arcsec$^2$ for the $r$-band image (on scales of 10 arcsec).

\subsection{Azimuthally averaged profiles} \label{profiles}

We explore the distribution of stellar light in the galaxy using azimuthally averaged surface brightness profiles. The profiles were extracted using the \texttt{photutils.isophote} package \citep{photutils}. The isophote-fitting routine automatically imposed a fixed position angle and ellipticity for isophotes with semi-major axes greater than 780 arcsec. The $r$-band isophotal solution was imposed on the $g$-band to ensure consistency. 

We measured the median sky inside a circular annulus centered on NGC 4565, with an inner radius of 1000 arcsec and an outer radius of 1125 arcsec (a width of 50 pixels). We considered the greatest difference between the median sky measured in annuli 50 pixels wide immediately inside and outside the sky annulus as a primary source of systematic uncertainty, as the position of the sky annulus is arbitrary and there are small variations in the background on scales of $\sim$10 arcmin (with an amplitude of less than 0.2\% of the mean sky value). We note that there is some overlap between the major axis of our azimuthally averaged profiles and the sky annulus, but due to their ellipticity the galaxy isophotes have little overlap with the sky annulus\footnote{More specifically, the area of major axis overlap with the sky annulus is small compared to the full annulus area, and measuring a median sky value instead of an average sky reduces the impact of a small population of somewhat brighter pixels.}.

To mitigate large scale, low amplitude variations in the background, the image was divided into four quadrants defined by the major and minor axes of the galaxy. (In particular, the southern half of the image was seen to be noticeably brighter than the northern half.) The $r$-band isophotal solution was applied to each quadrant in turn, and the sky was measured in each individual quadrant. The four quadrant profiles were weighted according to the number of unmasked pixels at each radius and averaged. The resulting combined profiles are shown in Figure~\ref{fig:profiles}, with systematics introduced by sky subtraction shown as shaded regions.

 \begin{figure*}[tb]
\plotone{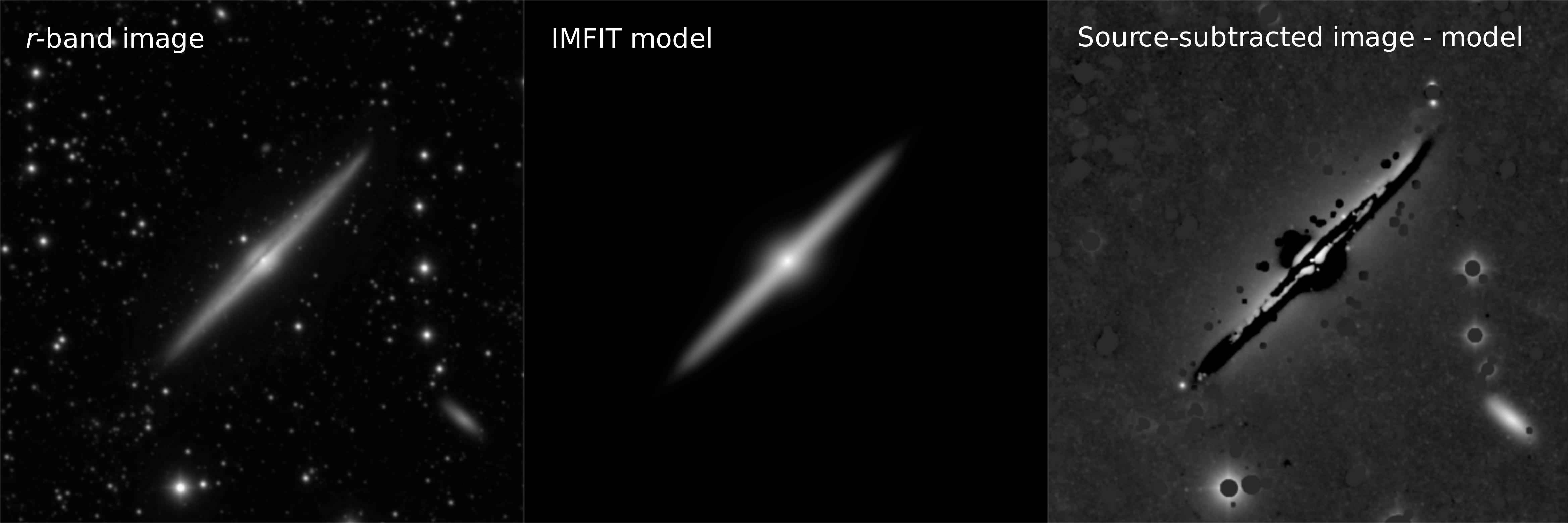}
\caption{ 2D $r$-band model fitted with IMFIT (centre panel), shown alongside the $r$-band image (left panel). The model consists of a \texttt{BrokenExponentialDisk3D} and a \texttt{Sersic} component. The right panel shows the residual image when the fitted model is subtracted from the source-subtracted $r$-band image.} 
\label{fig:2Dmodel}
\end{figure*}

We used IMFIT \citep{erwin15} to model the light distribution of NGC 4565, in order to assess the impact of the PSF on our profiles. The galaxy was fitted with a nearly edge-on broken exponential disc and a Sersic bulge. The darkest portions of the dust lane were initially masked during fitting to improve model convergence. The dust lane was then unmasked and the best-fit parameters from the first model were used as initial values for a second modelling run. For this second run, all parameters were held fixed except for intensities. Once the model parameters were finalized, model images were produced with and without PSF convolution. The $r$-band model image (produced with PSF convolution) and residuals are shown alongside the observed image in Figure~\ref{fig:2Dmodel}. 

Azimuthally averaged profiles were extracted from model images with and without PSF convolution, using the elliptical isophotes previously fitted to the $r$-band image. The resulting profiles are plotted alongside the $r$- and $g$-band profiles in Figure~\ref{fig:profiles}.  Since the kink in the observed profiles near 650 arcsec also appears in the model profiles but does not correspond to a model component or feature, this feature is at least partly explained by the isophote geometry. Indeed, there is a sudden drop in fitted ellipticity at this radius.

The profile of the PSF-convolved model image first becomes twice as bright as the non-PSF convolved model image at $472 \pm 12$ arcsec for $g$-band and $496 \pm 12$ arcsec for $r$-band. However, the observed profiles are much brighter than the model image profiles at these radii. If the difference between the PSF-convolved and non-PSF convolved profiles is taken as an estimate of the scattered light present at a given isophotal contour, only 8.6\% of the observed $r$-band profile and 13\% of the observed $g$-band profile can be explained by scattered light at $472 \pm 12$ arcsec. We repeated this test at an arbitrary position in the outer profile, $980 \pm 25$ arcsec. Here the estimated contribution from scattered light is even less for the $r$-band profile (3.2\%) and somewhat larger for the $g$-band profile (21\%). The estimated contribution of scattered light from the bright regions of the galaxy is not negligible (and may be somewhat underestimated in the outer profile; see the caveats below) but is far from sufficient to explain all of the light in the outskirts of the galaxy.

There are some important caveats to our 2D model comparison. The wide-angle PSF was only fitted out to a radius of 1200 arcsec, and the PSF cutout images used for model fitting and convolution were only 1250 x 1250 arcsec (radial scale of 625 arcsec). The accepted rule of thumb for the minor axis of edge-on galaxies is to measure the PSF on scales 1.5 times larger than the largest radius of interest \citep{sandin14}. This rule is relaxed to 1.1 times the largest radius of interest for early type galaxies and face-on discs \citep{sandin15}. We have not met this standard for the full extent of the area covered by our azimuthally averaged profiles, but the range of scales is sufficient for the scope of this paper. In the future, we will extend our PSF fitting range and increase the size of the PSF cutout images used for model fitting and convolution. This will be necessary for work measuring the stellar halo fraction of the galaxy. In the meantime we focus on the outskirts of the disc, which are less sensitive to the PSF and well-covered by our current PSF models.  

Due to the ellipticity of our outer isophotes, the effective radial extent of the profile in Figure~\ref{fig:profiles} is much smaller towards the minor axis. Furthermore, the galaxy is oriented at a $44^\circ$ angle to the pixel grid, which gives a factor of $\sim\sqrt{2}$ increase to the effective radial coverage of the PSF cutout along the major and minor axes of the galaxy. The resulting effective coverage of 883 arcsec is sufficient to estimate PSF contamination to a radius of $\sim$800 arcsec along the major axis and, conservatively, $\sim$590 arcsec along the minor axis (which would correspond to a semi-major axis length of $\sim$1140 arcsec in our azimuthally averaged profiles).

\subsection{Disc slice profiles} \label{slice_profiles}

Major axis profiles were extracted in thin slices extending from the center of NGC 4565 to the northwest and southeast. The position angle adopted was 134 degrees (East of North). We adopt a slice width of 11 pixels (27.5 arcsec or 1.7 kpc) to increase the S/N in the faint outer regions while still focusing on the disc midplane in bright regions. The procedure for measuring the sky was the same as for the azimuthally averaged profiles, though the quadrants were rotated 45 degrees with respect to the previous quadrants such that either disc limb fell in the center of a quadrant rather than lying at the boundary of two quadrants. 

\begin{figure*}[p]
\plotone{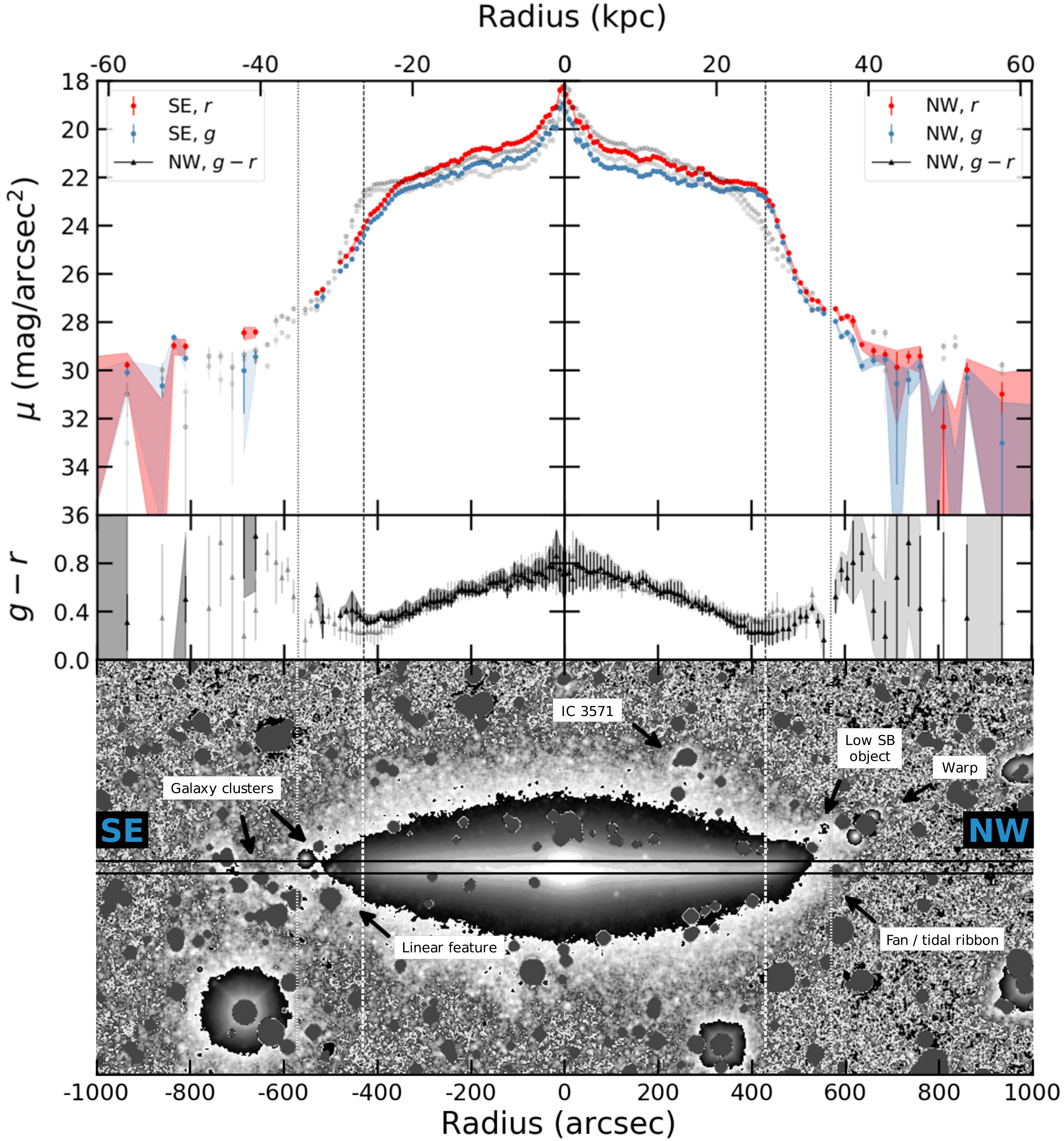}
\caption{ Southeast (left) and northwest (right) major axis $g$- and $r$-band profiles and colour profiles. On each side of this figure, the corresponding profiles from the opposite side of NGC 4565 are shown in faint grey for comparison. Error bars depict random errors in each bin and shaded error envelopes include systematic uncertainty in the measurement of the sky. The error bars in $g-r$ are relatively large for the inner regions due to the inclusion of two distinct populations of pixels within a bin (dusty and non-dusty) rather than low S/N in this region. The radii marked with vertical dashed and dotted lines are the approximate positions of the truncation and a post-truncation excess in the NW profiles, respectively. A rotated$r$-band image of NGC 4565 (with a sawtooth colour palette and Gauss histogram equalization scaling) is shown below the profiles with a common horizontal scale. The positions of several relevant features and objects are noted. The slice in which the major axis profiles were extracted is indicated by the two horizontal black lines spanning the width of the galaxy image.}
\label{fig:slice_profiles}
\end{figure*}

\begin{figure}[tb]
\plotone{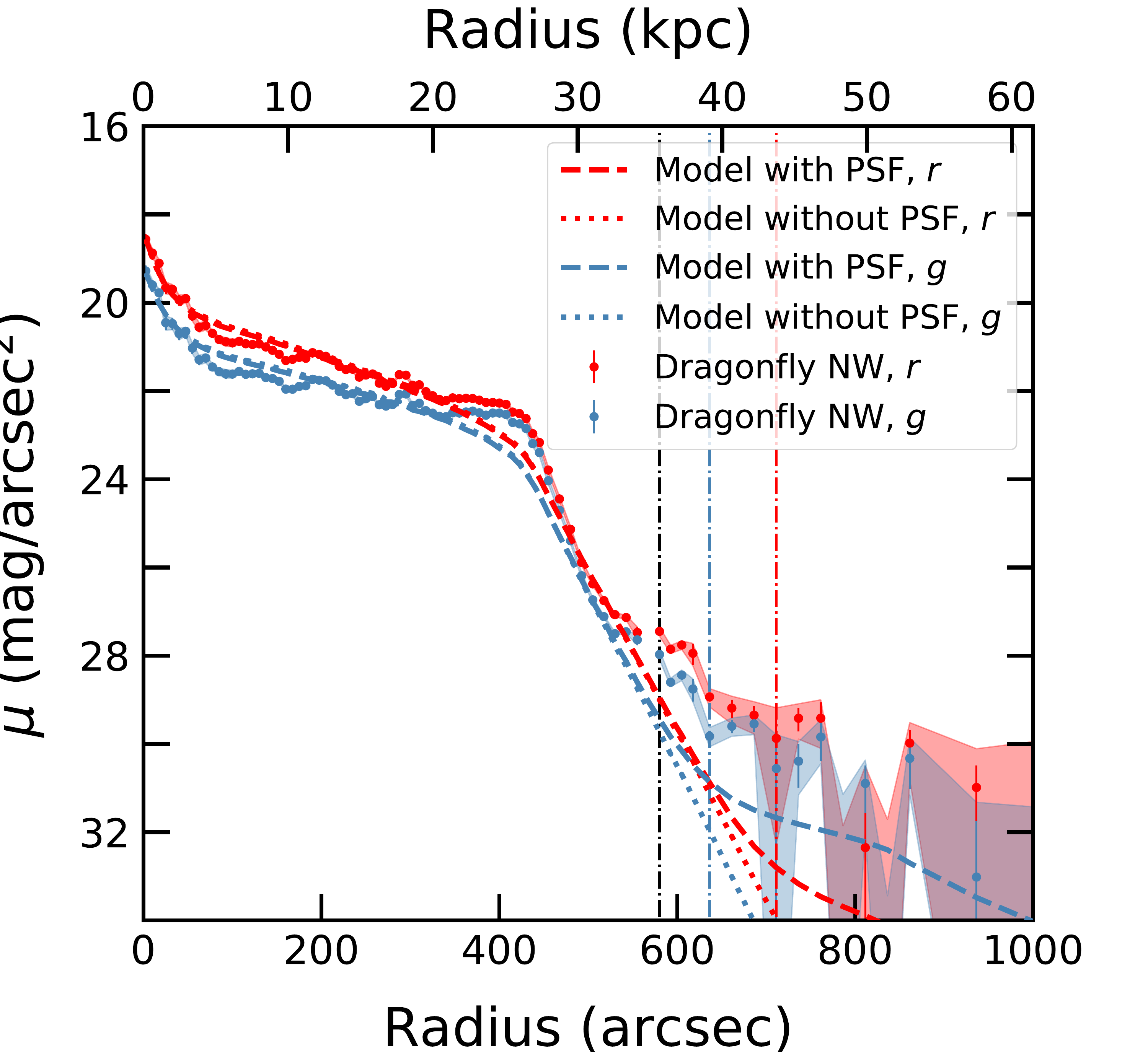}
\caption{Comparison of the $r$- and $g$- band NW major axis profiles with analogous profiles extracted from model images with and without PSF convolution. The red (blue) vertical dash-dotted line indicates the radius where the PSF-convolved $r$-band ($g$-band) model profile first becomes at least twice as bright as the model profile without PSF convolution. The black vertical dash-dotted line indicates the position in the vicinity of the fan where the contribution of scattered light to both observed profiles is estimated.} 
\label{fig:slice_PSF_test}
\end{figure}

We show the northwest (NW) and southeast (SE) disc slice surface brightness and colour profiles in Figure~\ref{fig:slice_profiles}. On each side of the disc, the profile from the opposite side is shown in greyscale to highlight their differences. Below the profiles, a rotated image of NGC 4565 is included for reference. The NW $r$-band ($g$-band) profile reaches a depth of $29.4^{+0.4}_{-0.5}$ mag/arcsec$^2$ ($29.8^{+0.4}_{-0.5}$ mag/arcsec$^2$) at a radius of $761 \pm 12$ arcsec, or $46.8 \pm 0.7$ kpc. The outer SE profiles are sparse due to heavy masking of background galaxy clusters, but do reach depths of $29.0^{+0.14}_{-0.16}$ mag/arcsec$^2$ and $29.5^{+0.12}_{-0.13}$ mag/arcsec$^2$ for the $r$- and $g$-band respectively at a radius of $811 \pm 12$ arcsec (or $49.9 \pm 0.7$ kpc). The impact of systematic sky uncertainty is generally large beyond this radius, and the point-to-point variation in the profiles becomes larger than the measurement errors.

We extracted major axis profiles from the model images with and without PSF convolution using the same technique as applied to the observed data. Figure~\ref{fig:slice_PSF_test} compares the resulting profiles to the NW $r$- and $g$-band profiles. The PSF-convolved profile first becomes twice as bright as the non-PSF convolved profile at $636 \pm 12$ arcsec for $g$-band and $711 \pm 12$ arcsec for $r$-band. The estimated contribution of scattered light to the observed profiles at these respective positions is 25\% and 34\% for the $g$-band profile, and 3.6\% and 4.4\% for the $r$-band profile. As with the azimuthally averaged profiles, scattered light from the bright inner regions of the galaxy is non-negligible in the outskirts but is not sufficient to account for most of the observed light. We reiterate that due to the limited size of the PSF cutout used for model fitting and convolution, we cannot confidently estimate the contribution of scattered light along the major axis beyond 800 arcsec. This is not a great concern as the point-to-point variation becomes larger than the measurement error in this regime, effectively marking the end of the reliable portion of the profile.

We confirm the overall disc asymmetry previously observed by \cite{naeslund97} and \cite{wu02}, and we find the asymmetry to be robust to changes in position angle of the NW and SE profile slices. This asymmetry is strongest near the disc truncation; the SE disc limb appears to truncate at a smaller radius, and much less sharply. The two sides become comparable again towards the end of the truncation regime, though there are large gaps in the SE profile in the post-truncation regime due to masking of background clusters and a star that was not deblended from NGC 4565 during source subtraction. The inner NW disc is somewhat dimmer than the inner SE disc. This may be a sign that material on the NW side has been redistributed towards the sharp truncation, or simply that there is more dust at the midplane on the NW side.

\subsubsection{Northwest disc limb} \label{NW_disc}

\begin{figure*}[tbp]
\plotone{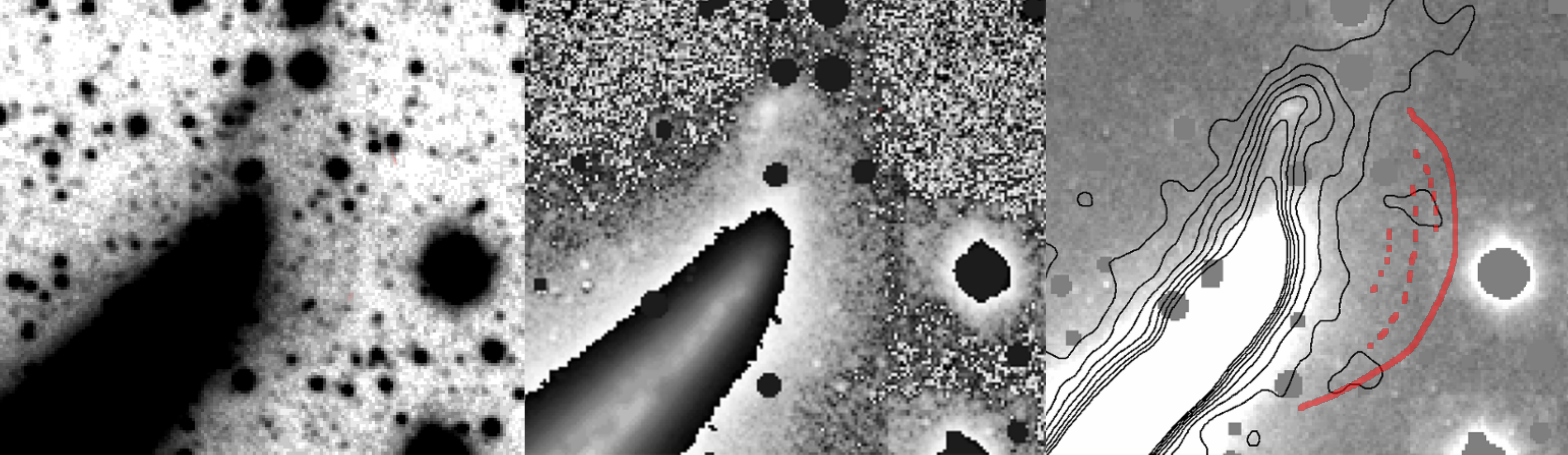}
\caption{Cut-outs at the NW edge of NGC 4565, without source subtraction (left), with source subtraction (center), and with WSRT HALOGAS HI contours superimposed \citep[right,][]{zschaechner12}. The approximate outer boundary of the fan is marked with a solid red line  and dashed red lines are used to indicate the position of ridges in the fan in the rightmost panel. Unsharp masking has been used on the left to enhance the fan-like shape extending to the right (west) of the bright disc plane. The fan seems to wrap around the NW disc edge. This feature is visible before and after source subtraction. Ridges are visible in the fan, though these may be due to coincidental positioning of compact sources.}
\label{fig:fan}
\end{figure*}

We trace the NW disc truncation over 8 kpc to a surface brightness of $\sim 27$~mag/arcsec$^\textrm{2}$, and confirm the end of the sharp truncation regime as first probed by \cite{wu02}. We find an extended excess of light in the post-truncation regime centred near 570 arcsec (35 kpc) that to our knowledge has not been previously reported. We mark this radius in Figure~\ref{fig:slice_profiles} with a dotted vertical line. A fan-like structure surrounding the NW edge of the disc is visible at the position corresponding to the excess light in the profile. This feature is highlighted in Figure~\ref{fig:fan}, and can also be seen in Figure~\ref{fig:slice_profiles}. The estimated contribution of scattered light to the major axis profiles at a radius of $580 \pm 6$ arcsec is only 6.1\% ($g$-band) and 1.1\% ($r$-band). We also note that the fan does not have a counterpart above the midplane. Therefore, we do not expect this feature to be induced by the PSF.

A few concentric ridges are visible in the fan. We highlight these features in Figure~\ref{fig:fan}. The ridges are visible in our images before and after source subtraction. It is unclear if these ridges are true features or the coincidental alignment of compact sources that lead the eye. Closer inspection of the CFHT MegaPipe image does show some low surface brightness objects below the detection threshold for MRF that line up with the ridges in the Dragonfly images.
 
Due to the depth of our imaging, we are able to measure major axis slice colour profiles through and beyond the truncation. The profile takes on a distinct u-shape about the NW truncation radius, reaching a minimum of $0.22 \pm 0.10$ at a radius of $430 \pm 4$ arcsec. Beyond the truncation the colour reddens again, except in the vicinity of the fan. Instead the colour becomes bluer, with a minimum lower than that reached about the truncation ($0.17 \pm 0.17$) at a radius of $555 \pm 6$ arcsec. This sharp decrease in $g-r$ occurs immediately before a small gap in the profiles due to automatically masked residuals from source subtraction, and therefore may be caused by contamination. Colour measurements in eight patches across the fan, south of the major axis profile footprint (avoiding subtracted sources and the ridges), range from 0.42 - 0.58. These values are consistent with the colour before and after the sharp blueward drop. Therefore, we do not consider this feature in the major axis $g-r$ profile to be representative of the fan's colour.

\subsubsection{Southeast disc limb} \label{SW_disc}

The transition from bright inner regions to outer disc is much smoother on the SE side of the disc than on the NW side. Additionally, the break radius on the SE side is smaller than the NW side ($\sim$360 arcsec versus $430 \pm 4$ arcsec, or 22 kpc versus 26.4 kpc). The outer SE colour profile displays a u-shape similar to that in the NW colour profile. The colour remains relatively constant between 360 and 430 arcsec (reaching a minimum of $0.31 \pm 0.05$ at $422 \pm 4$ arcsec) and does not begin to redden again until beyond the \emph{NW truncation} radius. We note that the minimum $g-r$ colour on the SE side of the disc is redder than that on the NW side ($0.31 \pm 0.05$ versus $0.22 \pm 0.10$).

As discussed in Section~\ref{source_sub}, broad masking is required beyond $\sim$500 arcsec (31 kpc) due to background galaxy clusters and improperly subtracted stars. Some milder contamination may extend outside the masked region, so caution is required when considering the few unmasked data points in the outer SE profile. 

\subsubsection{Comparison with literature profiles}

\begin{figure*}[tbp]
\plotone{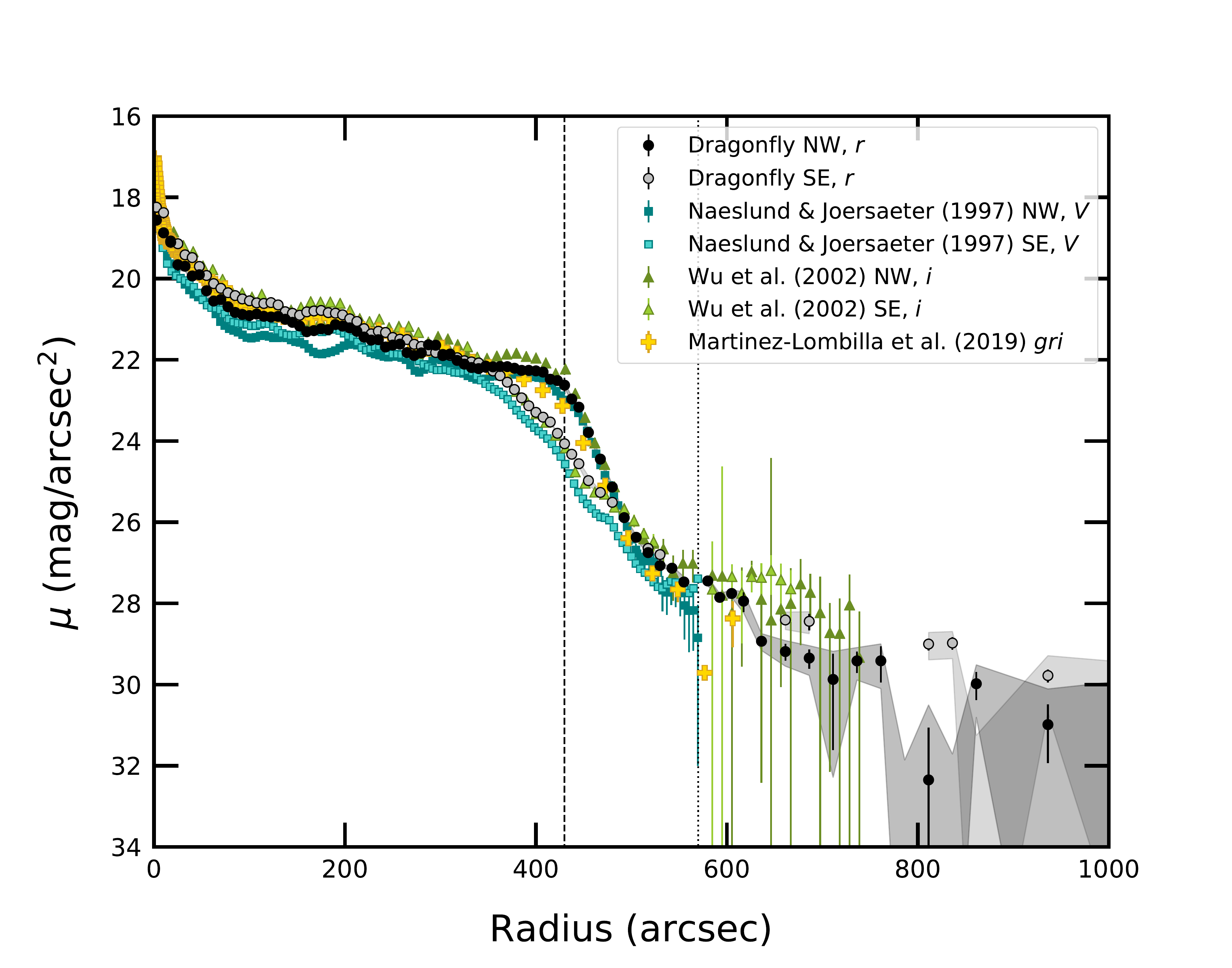}
\caption{Dragonfly NW and SE $r$-band profiles alongside previously published major axis profiles. Point shape and colour is used to indicate profiles from the same work, with a darker colour for the NW profile and a lighter colour for the SE profile. The exception is the profile from \cite{martinezlombilla19}, shown with yellow crosses, which is an average of the NW and SE profiles.}
\label{fig:compare}
\end{figure*}

Figure~\ref{fig:compare} shows the major axis profiles produced in this work alongside previously published major axis profiles. There is good agreement across all work on the differing shapes of the NW and SE profiles and the position of the NW truncation. Agreement remains excellent through the truncation regime, to a radius of $\sim600$ arcsec (37 kpc). Beyond this radius, our NW profile extends farther than other work with smaller error bars. \cite{wu02} observed a shallow outer component (which they modelled as a power law stellar halo), but our profiles continue to decrease more rapidly in this regime. The relative excess observed by \cite{wu02} may be attributable to scattered light; \cite{sandin15} demonstrated that a similar upturn in minor axis profiles of NGC 4565 can be explained by scattered light, but did not repeat the same test for major axis profiles. The sharp upturn towards the end of the \cite{naeslund97} SE profile could be contamination from previously noted background clusters.

There is broad agreement between our NW profiles and the GHOSTS NW major axis stellar density profile \citep{harmsen17}. The inflection point marking the end of the first truncation is not discernible in their profile  (see their Figure~6) due to a gap between HST pointings and larger radial bins, but their outer points are suggestive of an outer component with a longer scalelength.

\section{Discussion} \label{discussion}

\begin{figure*}[tb]
\plotone{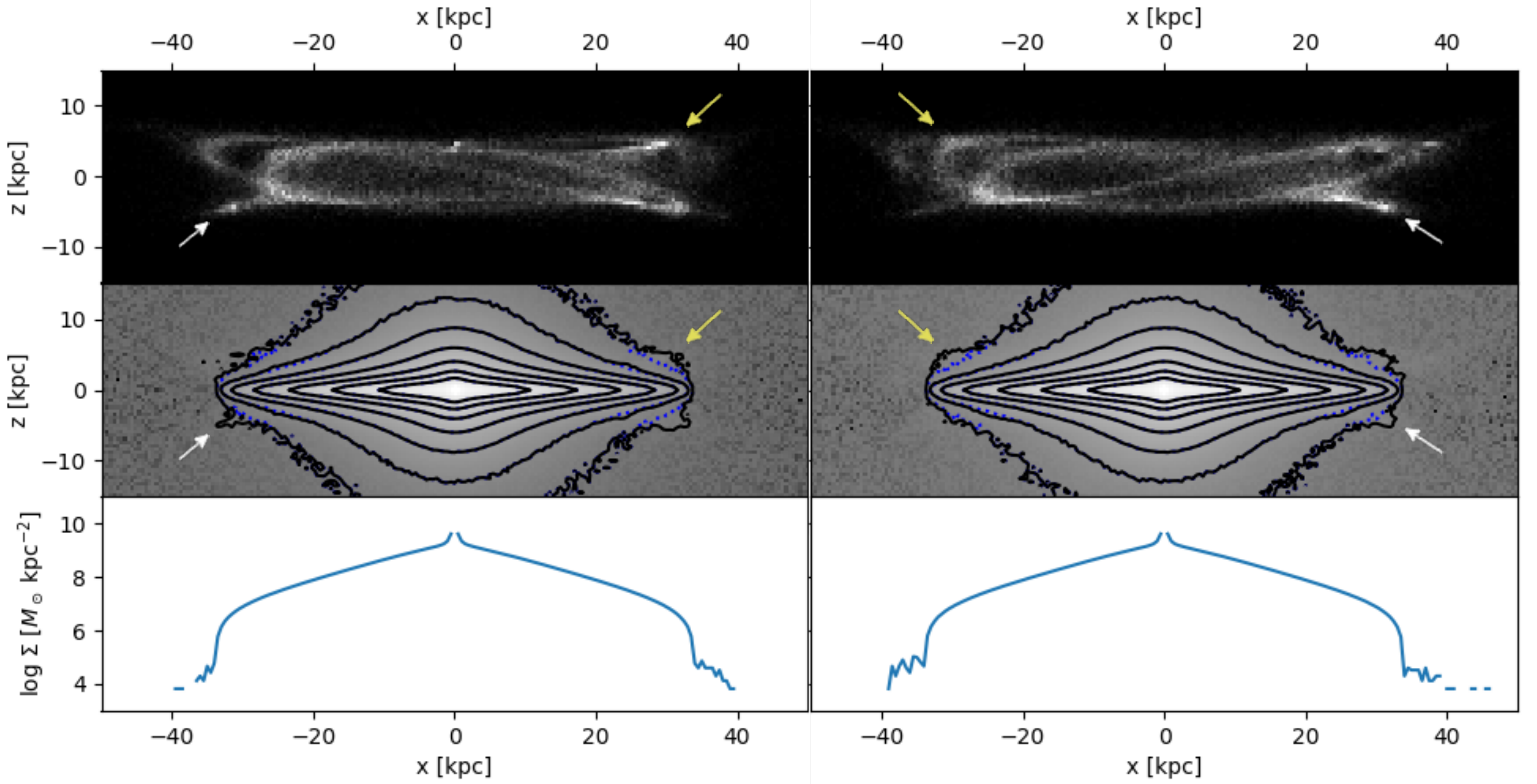}
\caption{Two example tidal ribbons viewed in the x-z plane. The top panels show the tidal ribbon mass density in projection. The middle panels show the sum of the tidal ribbon and the simulated galaxy surface mass densities, representing a mock image of the system when viewed edge-on. Noise has been added to the mock images for display purposes. Any negative mass densities (outside the region where three Miyamoto Nagai discs closely approximate an exponential disc) are set to zero instead. The black contours span $10^5 M_\odot \textrm{ kpc}^{-2}$ to $10^8 M_\odot \textrm{ kpc}^{-2}$ in half order-of-magnitude increments. The dotted blue contours trace only the simulated galaxy, and so the differences between the blue and black contours indicate additional material contributed by the tidal ribbons. Yellow arrows indicate the fan-like features induced in each galaxy due to the tidal ribbon. White arrows have also been included to indicate linear features. Mass density profiles through the midplane are shown in the bottom panels. All panels share a common horizontal axis. }
\label{fig:ribbons}
\end{figure*}

In order to grow stellar discs of the appropriate mass and radius at the present date, the accretion of satellite galaxies must have played a part \citep{bluck12, ownsworth12, sachdeva15}. \cite{karademir19} demonstrated that it is possible to significantly increase the size of a disc without altering the disc's scaleheight via mergers with small satellites (mass ratios of 1:50 or 1:100) when the satellite approaches at low inclination to the disc plane. Evidence of disc growth via accretion of externally formed stars can be found in cosmological hydrodynamic simulations \citep{pillepich15, rodriguezgomez16} and can be inferred from the ages and motions of stars in our own Galaxy \citep{quillen01, gomez12}. However, observing such evidence in other galaxies is challenging.

Recently, \cite{zhang18} detected an extended low surface brightness disc surrounding NGC 2841 out to distances of 70 kpc. Radial migration and in-situ star formation were both identified as likely contributors to the formation of the outer disc, but were unable to fully account for the disc's present day size and mass. A warp in the extended disc suggested accretion has also played an important role in the formation of this structure. 
 
Below, we discuss distinct features of NGC 4565's disc that suggest growth or evolution by accretion of satellite galaxies; namely, the galaxy's strong disc asymmetry and the northwestern fan. We consider plausible scenarios that may explain these features. 

\subsection{Evidence for accretion-based growth} \label{evidence}

The disc of NGC 4565 is strongly asymmetric through the midplane, with the strongest asymmetry near the NW truncation radius.  The new fan-like feature we have found on the NW side is another asymmetry between the two sides. Together with the one-sided HI warp, these asymmetries indicate that one or more events or processes have not had an equal influence across the disc. Such processes would likely have external causes rather than arising due to secular evolution. It is not immediately clear how the striking disc asymmetries of NGC 4565 were shaped. We do know that it is not unusual for galaxies to be at least weakly asymmetric, in stars as well as HI \citep{haynes98, vaneymeren11}.

\cite{roskar08} demonstrated that down-bending disc breaks may arise due to the combination of a radially-declining star formation rate and radial migration. A notable signature of such disc breaks is a u-shaped age gradient, where the minimum stellar age is reached at the break radius. We report a u-shaped $g-r$ gradient on the NW side of the disc, and a similar but weaker feature on the SE side of the disc. If we consider $g-r$ colour as an indicator of average stellar age, our observations of the galaxy's truncation and colour profile resemble the scenario presented by \cite{roskar08}. This would suggest that radial migration has built up the post-truncation disc.  However, this process cannot account for the strong disc asymmetry, the fan, or the shallower post-truncation disc. Another (or, at least an additional) explanation is needed.

The fan's well-defined shape (Figures~\ref{fig:slice_profiles} and \ref{fig:fan}) suggests a dynamically distinct structure. It seems to lie in the disc plane right at the disc's visual edge, wrapping around the edge of the disc with minimal extent above the midplane. There is a hint of the fan visible in the density of old stars plotted in Figure~2 of \cite{radburnsmith14}. The optical warp that follows the HI warp is traced by younger ($\lesssim 1$~Gyr) stars rather than older stars, and so the fan is not likely to be directly associated with the warp. While the fan does not appear in the \cite{harmsen17} major axis density profile of red giant branch stars, this is likely due to a combination of coarse binning and the gap at 35-45 kpc (see their Fig. 6). 

\subsubsection{Identifying progenitor candidates in the vicinity of the disc} \label{progenitors}

 We note a low surface brightness object coincident with the warp. Its position is marked in Figure~\ref{fig:slice_profiles}, and it can also be clearly seen in Figure~\ref{fig:fan}. Previous HST imaging has resolved this object into stars, and it was deemed to be a massive star cluster \citep{radburnsmith14}. An alternative interpretation is that this object may be the remnant of a recently accreted dwarf galaxy. If this is the case, it may be associated with the creation of the fan. 

NGC 4565 hosts surviving satellites which may be interacting or may have recently passed through the disc plane. An obvious candidate is IC 3571, the dwarf irregular galaxy just north of the disc (noted in Figure~\ref{fig:slice_profiles}). It has a projected separation of 348 arcsec (21 kpc) from the centre of NGC 4565 (240 arcsec or about 15 kpc perpendicular to the disc plane). It is connected to NGC 4565 in HI \citep{zschaechner12} and therefore is clearly interacting. Another candidate that stands out is NGC 4562 \citep[$M_* = 8.13 \times 10^{8} M_\odot$,][]{sheth10}, a late spiral 800 arcsec (49 kpc) southwest of the centre of NGC 4565 (noted in Figure~\ref{fig:quad_images}). 

\subsection{The fan as a tidal ribbon} \label{ribbon}

Recently, \cite{dehnen18} showed that the dynamical conditions necessary for nearly-one-dimensional stream formation can be strongly violated in the case of a satellite on a circular orbit sufficiently close to the disk plane. In this case the vertical oscillation frequency varies substantially between stars and so they are dispersed perpendicular to the satellite's orbit as well as along the orbit, forming a band or `tidal ribbon'.  We suggest that the NW fan may consist of ribbon-like debris from a dwarf satellite in the process of being disrupted by NGC 4565. This class of tidal debris has not, to our knowledge, been observed previously.

To investigate this possibility, we have produced numerical simulations of tidal ribbons that qualitatively match both the global properties of NGC 4565 and the appearance of its fan using the Lagrange-stripping method of \cite{2015MNRAS.452..301F}, as implemented in the galactic dynamics package \texttt{gala} \citep{gala}. For a given progenitor orbit and mass in a particular host potential, test particles are released at each time step with positions and velocities consistent with those expected for stars being stripped by tides and then integrated forward to the present. This technique produces realistic models quickly and allows a substantive exploration of parameter space.

The potential used in our simulations was constructed to approximately match NGC 4565's HI rotation curve. The disc was modelled with a triple Miyamoto-Nagai (MN) approximation to an exponential disc \citep{smith15}. The triple MN disc mass density deviates from exponential at large radii and eventually becomes negative, but our model has positive mass densities everywhere due to the dark matter halo. The total mass of the disc is $10^{11} M_\odot$ and the radial scale length is 5 kpc. The dark matter halo was represented by a $10^{12} M_\odot$ Navarro-Frenk-White profile \citep{nfw96} with $r_s = 25$ kpc. A Hernquist bulge \citep{hernquist90} of mass $3\times10^{10} M_\odot$ with a scale radius of 0.5 kpc was included as well. 

We generated many tidal ribbons with varying initial radius (30 - 36 kpc) and velocity in the plane (190 - 220 km/s), viewed from four distinct edge-on sightlines. After some experimentation, the initial vertical velocity and displacement were fixed at 20 km/s and 4 kpc to match the apparent width of the fan. Two representative tidal ribbons are shown in Figure~\ref{fig:ribbons}. We consider surface mass density ``images'' as a simple proxy for visible light images. Both mock images show excess mass contributed by the ribbon affecting the shape of the overall surface mass density distribution near the edge of the disc, but having minimal impact on the rest of the galaxy. From our imaging, we estimate the surface mass density of the ribbon surrounding NGC 4565 at $10^5 M_\odot \textrm{ kpc}^{-2}$ near the NW disc limb. The simulated ribbons' mass densities at the edge of the disc approximately matches this value.

We have shown the results of 1.5 Gyr of interaction, but we were able to achieve similar structures with 1 Gyr of interaction. For longer simulations, the debris becomes more and more evenly distributed perpendicular to the progenitor's orbit (more ``ribbon-like''). We have fixed the ribbon progenitor mass at $10^8 M_\odot$; we are able to produce similar ribbons for less and more massive progenitors, but the impact of the debris on mock galaxy images is most similar to NGC 4565 for this mass. 
Both mock images include at least one rounded lobe that seems to wrap around one edge of the disc. The first image has rounded lobes above and below the plane on the right side of the disc, while the second image shows a rounded lobe only above the plane on the left side of the disc. The low surface brightness disc outskirts north and east of NGC 4565's NW disc limb appear to drop off faster than anywhere else around the galaxy (Figure~\ref{fig:slice_profiles}), and so the latter tidal ribbon morphology/orientation is a better match to this galaxy. 

The simulated tidal ribbons both introduce new features at either side of their host galaxy's disc. So far, we have focused on the prominent features at the NW disc limb and have neglected the opposite side of the disc. There does seem to be a linear feature extending from the south side of the disc towards a bright (masked) star, noted in Figure~\ref{fig:slice_profiles}. Similar linear features are seen in the contours of the mock images with tidal ribbons. The low surface brightness glow surrounding the disc also seems to be more puffy north of the SE disc limb.

\subsection{Other scenarios to consider} \label{other}

The tidal ribbon explanation is not unique, and does not preclude other processes from growing or influencing the outer disc of NGC 4565. Minor mergers \citep{zaritsky97}, tidal interactions \citep{kornreich02}, and accretion of gas from the IGM \citep{bournaud05} have all been suggested as possible causes of asymmetry in spiral galaxies. Halo asymmetry or outer disc instabilities have also been identified as possible internal causes of asymmetry \citep{zaritsky13}. In the case of NGC 4565, an external origin would not be surprising given the galaxy's asymmetric warp at large radii (present in young stars and in HI).

The stronger NW truncation can be viewed as an excess relative to the SE side of the disc. In this case, the bluer colour about the truncation could be indicative of a recent gas-rich accretion event that has deposited material primarily at the truncation radius. Qualitatively, the bluest regions of the NW colour profile fall in the radial range where the NW surface brightness exceeds the SE surface brightness. Outside of this range, the colours of both sides of the disc are more similar. In this interpretation, the underlying disc of NGC 4565 is approximately symmetric in surface brightness and colour and there is little change in outer disc colour from 25 - 34 kpc. 

The fan and disc truncation asymmetry may have been induced by a satellite flyby. \cite{pricewhelan15} demonstrated that a satellite galaxy passing vertically through the midplane at a large distance of 80 kpc can induce asymmetric rings in the disc around 20-40 kpc. These rings propagated outwards and oscillated above and below the midplane by a few kpc, similar to the amplitude of the fan (see their Figure~8). 

While the optical warp and the fan appear to be separate features due to their differing stellar populations, they may share a common origin. The stellar content of the warp region implies a recent gas-rich accretion event \citep{radburnsmith14}. The fan could consist of disc stars ejected or excited by this event. Similarly, the fan could be associated with the disc's asymmetry if it may be explained by another accretion event or by a satellite fly-by. The outermost HI contours in the right panel of Figure~\ref{fig:fan} are similar to the fan's shape, suggesting that there may be a gaseous component to the fan and that both disc stars and gas were ejected by the perturbing event.

\section{Summary} \label{summary}

We have presented deep imaging of NGC 4565, one of the first galaxies from the Dragonfly Edge-on Galaxies Survey to be observed and reduced. In major axis slice profiles targeting the disc, we have found an end to the sharp NW disc truncation, a fan-like feature in the post-truncation disc, and another outer inflection point in the NW profile. We identified a distinct u-shape in the colour profiles about the radius of the NW truncation. 

Several possible origins for the disc asymmetry and fan feature were identified. We gave particular attention to the tidal ribbon hypothesis and produced some simple simulations that qualitatively replicate the fan. All of the scenarios outlined that may have helped shape the features seen on the NW side of NGC 4565 require some interaction with a satellite galaxy. We identified what may be the remnant of a dwarf galaxy in the vicinity of the warp, and IC 3571 and NGC 4562 are also two candidate perturbers. There are several other dwarf galaxies with similar line-of-sight velocities to NGC 4565 that are likely satellites. Additionally, many low surface brightness satellite candidates have been identified in our imaging (S. Danieli et al., in prep). 

We drew a connection with the recently discovered extended low surface brightness disc surrounding NGC 2841 \citep{zhang18} as another feature that appears to have at least a partially accreted origin. The existence of these structures may inform the distribution of satellite galaxies around their hosts (and similar systems) as some formation scenarios require specific orbit parameters and orientations. To form a tidal ribbon, the satellite must be accreted in the plane on a disc-like orbit (with vertical oscillations comparable to the disc scale height). Any growth via accretion of the extended low surface brightness disc surrounding NGC 2841 would also require satellites with similar angular momentum to the existing stellar material \citep{zhang18}. Similarly, simulations of ``mini-mergers'' in the disc plane yielded radial disc growth without increasing the disc's scaleheight \citep{karademir19}.

If satellite galaxies are generally accreted isotropically, low surface brightness structures like these in the outskirts of discs should be rare. If instead they are common, this would suggest a preferred direction for accretion of satellites. Our ongoing study of edge-on spiral galaxies through DEGS (C. Gilhuly et al., in prep) will help clarify the frequency of these structures.  The role that the accretion of satellite galaxies plays in (thin) disc growth merits greater attention. 
 
\acknowledgments

We thank all of the staff at New Mexico Skies for their crucial role in Dragonfly operations. We also thank L. Zschaechner for providing a zeroth moment HI map of NGC 4565, and M. N\"{a}slund and C. Mart\'inez-Lombilla for sharing their major axis profiles of NGC 4565 in machine-readable formats. We are grateful to the anonymous referee for their comments and suggestions which have greatly improved this paper. Support for this work was provided by NSERC, the Dunlap Institute, and MPIA. DH acknowledges financial support from the Natural Sciences and Engineering Research Council of Canada (NSERC; funding reference number RGPIN-2015-05235) and an Ontario Early Researcher Award (ER16-12-061). This research has made use of the NASA/IPAC Extragalactic Database (NED), which is operated by the Jet Propulsion Laboratory, California Institute of Technology, under contract with the National Aeronautics and Space Administration. This research has also made use of NASA's Astrophysics Data System Bibliographic Services. The authors have enjoyed using this website for conversion between angular and physical scales: \url{http://arcsec2parsec.joseonorbe.com/index.html}

\bibliographystyle{aasjournal}
\bibliography{refs.bib}

\end{document}